\documentclass[letterpaper,twocolumn,american,showpacs,pre,aps,superscriptaddress,eqsecnum]{revtex4}
\usepackage[latin1]{inputenc}
\usepackage{bm}
\usepackage{multirow,amssymb,amsbsy,amsmath}
\usepackage{stmaryrd}
\usepackage{graphicx}
\usepackage{hyperref}
\makeatletter
\usepackage{pifont}
\makeatother

\newcommand*{\Haver}[1]{\mathopen{\llbracket} #1 \mathclose{\rrbracket}}
\newcommand*{\bra}[1]{\mathopen{\langle}#1\mathclose{|}}
\newcommand*{\ket}[1]{\mathopen{|}#1\mathclose{\rangle}}

\begin{document}

\title{Non-Markovian quantum dynamics: Correlated
       projection superoperators and Hilbert space averaging}

\author{Heinz-Peter Breuer}

\email{breuer@physik.uni-freiburg.de}

\affiliation{Physikalisches Institut, Universit\"at Freiburg,
             Hermann-Herder-Str.~3, D-79104 Freiburg, Germany}

\author{Jochen Gemmer}

\email{jgemmer@uos.de}

\affiliation{Fachbereich Physik, Universit\"at Osnabr\"uck,
             Barbarastr.~7, D-49069 Osnabr\"uck, Germany}

\author{Mathias Michel}

\email{mathias.michel@itp1.uni-stuttgart.de}

\affiliation{Institut f\"ur Theoretische Physik, Universit\"at
             Stuttgart, Pfaffenwaldring 57, D-70550 Stuttgart, Germany}

\date{\today}

\begin{abstract}
The time-convolutionless (TCL) projection operator technique
allows a systematic analysis of the non-Markovian quantum dynamics
of open systems. We introduce a class of projection superoperators
which project the states of the total system onto certain
correlated system-environment states. It is shown that the
application of the TCL technique to this class of correlated
superoperators enables the non-perturbative treatment of the
dynamics of system-environment models for which the standard
approach fails in any finite order of the coupling strength. We
demonstrate further that the correlated superoperators correspond
to the idea of a best guess of conditional quantum expectations
which is determined by a suitable Hilbert space average. The
general approach is illustrated by means of the model of a spin
which interacts through randomly distributed couplings with a
finite reservoir consisting of two energy bands. Extensive
numerical simulations of the full Schr\"odinger equation of the
model reveal the power and efficiency of the method.
\end{abstract}


\maketitle

\section{Introduction}
Realistic quantum mechanical systems are influenced through the
coupling to an environment which contains a large number of mostly
uncontrollable degrees of freedom. The unavoidable interaction of
an open quantum systems with its environment gives rise to the
mechanisms of damping and dissipation, and to a strong and often
rapid loss of quantum coherence. Applications of the theory of
open quantum systems \cite{TheWork} are found in almost all areas
of physics, ranging from quantum optics \cite{WallsMilburn} to
condensed matter physics \cite{Weiss} and chemical physics
\cite{Pechukas}, from quantum information \cite{Nielsen} to
spintronics \cite{Loss}. Moreover, the theory of open quantum
systems provides the foundations of quantum measurement theory
\cite{Braginsky}, decoherence \cite{Giulini} and the emergence of
thermodynamic behavior \cite{GEMMER}.

In a microscopic approach one regards the total system, which is
composed of the open system $S$ and its environment $E$, as a
closed quantum system following a Hamiltonian time evolution. One
of the central goals of the theoretical treatment is then the
analysis of the dynamical behavior of the populations and
coherences which are given by the elements of the reduced density
matrix $\rho_S(t)={\mathrm{tr}}_E\rho(t)$. Here, $\rho(t)$ denotes
the density matrix of the composite system and ${\mathrm{tr}}_E$
the partial trace taken over the environment.

In the Markovian regime a complete mathematical theory is
available which is based on the concepts of completely positive
quantum dynamical semigroups and the corresponding Markovian
master equations in Lindblad form
\cite{GORINI1,GORINI2,LINDBLAD,SPOHN}. However, if the physical
conditions underlying the Markov approximation are violated one
has to cope with strong non-perturbative and memory effects and
the theoretical and mathematical treatment of the reduced system
dynamics is typically much more involved.

A systematic approach to non-Markovian dynamics is provided by the
projection operator techniques \cite{NAKAJIMA,ZWANZIG} which are
extensively used in nonequilibrium thermodynamics and statistical
mechanics \cite{GRABERT,KUBO}. The key concept of these techniques
consists in the introduction of a certain projection superoperator
${\mathcal{P}}$ which acts on the operators of the state space of
the total system. The superoperator ${\mathcal{P}}$ formalizes the
idea of the elimination of degrees of freedom from the complete
description of the states of the total system. Thus, if $\rho$ is
the density matrix of the composite system, the projection
${\mathcal{P}}\rho$ serves to represent a simplified effective
description through a reduced set of variables. For this reason,
the projection ${\mathcal{P}}\rho$ is called the {\textit{relevant
part}} of the total density matrix, while the complementary
projection ${\mathcal{Q}}\rho=\rho-{\mathcal{P}}\rho$ is referred
to as {\textit{irrelevant part}}.

With the help of the projection operator techniques one derives
closed equations of motion for the relevant part
${\mathcal{P}}\rho(t)$ from which one obtains approximate master
equations by means of a systematic perturbation expansion with
respect to the system-environment coupling. We shall concentrate
in this paper on a specific variant of the technique which is
known as the time-convolutionless (TCL) projection operator method
\cite{SHIBATA1,SHIBATA2,SHIBATA3,SHIBATA4,ROYER}. The advantage of
this formulation is that it leads to dynamic equations for
${\mathcal{P}}\rho(t)$ which are local in time and involve an
explicitly time-dependent generator. A general account of
projection operator methods and, in particular, of the TCL
approach and its applications may be found in Ref.~\cite{TheWork}.

In the standard approach to the dynamics of open systems one
chooses a projection superoperator which is defined by the
expression ${\mathcal{P}}\rho = \rho_S \otimes \rho_E$, where
$\rho_E$ is some fixed environmental state. This superoperator
projects the total state $\rho$ onto an un-correlated tensor
product state. Since $\rho_E$ is considered as fixed, it implies
that the elements of the reduced density matrix $\rho_S(t)$
represent the relevant variables used for an effective description
of the reduced system dynamics. This ansatz for the projection
superoperator ${\mathcal{P}}$ is widely used in studies of open
quantum systems. It has been used to derive Markovian and
non-Markovian quantum master equations for many applications (see,
e.~g., \cite{CHANG,DESPOSITO,ANNPHYS,ATOM-LASER,BKP}). Moreover,
non-Markovian generalized master equations for the reduced density
matrix have been developed on the basis of phenomenological
considerations \cite{LIDAR}.

The paradigm of these and many other approaches is the usage of
the reduced density matrix $\rho_S(t)$ as dynamical variable for
which appropriate (exact or approximate) dynamic equations are to
be developed. However, it is important to realize that the
projection operator techniques are much more general and flexible,
and that they offer many further possibilities for the
construction of suitable projection superoperators. The only
formal condition which must be satisfied in order to apply the
techniques is that ${\mathcal{P}}$ is a map which operates on the
total state space and has the property of a projection operator,
i.~e., ${\mathcal{P}^2=\mathcal{P}}$. This is a very general
condition which can be fulfilled in many different ways.

We mention two examples: In the analysis of classical stochastic
processes one considers a projection of the form
${\mathcal{P}}X(t) = \langle X(t) \rangle$ which takes any
stochastic process $X(t)$ to its average $\langle X(t) \rangle$
\cite{SHIBATA1}. With this choice the TCL technique leads to a
cumulant expansion of the dynamic equation for the average
\cite{KUBO63,ROYER72,KAMPEN1,KAMPEN2}. In nonequilibrium
thermodynamics a further projection of the form
${\mathcal{P}}\rho=\rho_{\mathrm{diag}}$ is often introduced which
maps any density matrix $\rho$ to its diagonal part
$\rho_{\mathrm{diag}}$ in a suitably chosen basis, yielding the
famous Pauli master equation in lowest order of the TCL expansion
\cite{KUBO}.

In the present paper we shall construct a class of projection
superoperators ${\mathcal{P}}$ which enable the non-perturbative
treatment of highly non-Markovian processes in open quantum
systems. These superoperators project the state of the total
system onto a correlated system-environment state, i.~e., onto a
state which contains statistical correlations between certain
system and environment states. Thus, we give up the paradigm of
using the reduced density matrix as the dynamical variable, and
enlarge the set of relevant variables to account for statistical
correlations which are responsible for strong non-Markovian
effects. The idea of introducing additional variables has been
realized in different ways and used in various contexts
\cite{GASPARD,GARRAWAY1,GARRAWAY2,IMAMOGLU,GQT,BUDINI}. Here, we
implement this idea directly in the definition of a projection
superoperator and connect it with the method of the TCL technique.
This connection enables us to determine higher order corrections
in a systematic way and, hence, to assess the quality of the
approximations obtained.

Recently, an entirely different approach has been suggested, the
Hilbert space average method (HAM) \cite{GEMMER,Gemmer01}. This method
employs the concept of a {\textit{best guess}} for conditional
quantum expectation values. It is based on the determination of a
conditional Hilbert space average. The method provides us with a
systematic principle to estimate quantum expectation values
conditioned on prescribed values for the expectations of a given
set of operators on the total state space. HAM can be used to
construct effective equations of motion for the given set of
operators and, hence, yields an alternative approach to
non-Markovian quantum dynamics. It will be shown here that the
method of the Hilbert space average and the projection operator
techniques which are based on the class of correlated projection
superoperators are closely related. In fact, we are going to
demonstrate that HAM represents the lowest order of the TCL
expansion corresponding to this class of superoperators.

The application and the efficiency of our approach will be
illustrated and discussed here by means of a specific
system-reservoir model. The model consists of a spin which
interacts with two environmental energy bands through a set of
random couplings \cite{Gemmer2005,Gemmer2005a}. This model
exhibits an unexpected feature. Namely, it turns out that the
usual Born-Markov approximation fails for this model, although the
standard Markov condition is satisfied, i.~e., although the width
of the environmental two-point correlation function is small
compared to the relaxation time. By contrast, it will be
demonstrated by means of a comparison with the numerical solution
of the full Schr\"odinger equation of the model that the TCL
expansion using the correlated projection superoperator yields
accurate results already in lowest order of the perturbation
expansion.

The paper is organized as follows. The class of correlated
projections will be introduced in Sec.~\ref{Sec-Projection}. This
section also contains a brief general account of the projection
operator techniques, and introduces the basic dynamic equations as
well as the perturbation expansion of the master equations for the
relevant variables. The principles and equations of the Hilbert
space averaging method are outlined in Sec.~\ref{HAM-RED}, where
we will also describe the connection between the principles of HAM
and the structure of the correlated projection superoperators.
Sec.~\ref{Sec-Appl} contains the application of the general
concepts developed here to a specific system-reservoir model. We
shall discuss in detail the origin of the failure of the
Born-Markov approximation, and show that and why the new
projection superoperators yield an efficient and accurate
approximation of the dynamics. Finally, we draw our conclusions in
Sec.~\ref{CONCLU}.

\section{Projection operator techniques}\label{Sec-Projection}

\subsection{Projection superoperators}\label{Sec-Superoperators}
We consider an open quantum system $S$ with state space
${\mathcal{H}}_S$ which is coupled to an environment $E$ with
state space ${\mathcal{H}}_E$. The Hilbert space of states of the
composite system is given by the tensor product
${\mathcal{H}}={\mathcal{H}}_S\otimes{\mathcal{H}}_E$. We assume
that the dynamics of the total density matrix $\rho(t)$ of the
composite system is governed by some Hamiltonian of the form
$H=H_0+V$, where $H_0$ generates the free time evolution of the
system and of the environment, and $V$ describes the
system-environment coupling. We work in the interaction
representation and write the von Neumann equation of the combined
system as
\begin{equation} \label{NEUMANN}
 \frac{d}{dt} \rho(t)
 = -i [V(t), \rho(t) ] \equiv {\mathcal{L}}(t)
 \rho(t),
\end{equation}
where $V(t)$ is the Hamiltonian in the interaction picture, and
${\mathcal{L}}(t)$ denotes the corresponding Liouville
superoperator.

The projection operator techniques are based on the introduction
of a projection superoperator ${\mathcal{P}}$. This is a linear
map
\begin{equation}
 \rho \mapsto {\mathcal{P}}\rho
\end{equation}
which takes any operator $\rho$ on the total state space
${\mathcal{H}}$ to an operator ${\mathcal{P}}\rho$ on
${\mathcal{H}}$, and which has the property of a projection
operator:
\begin{equation} \label{PROJECTION}
 {\mathcal{P}}^2 = {\mathcal{P}}.
\end{equation}
Given a map ${\mathcal{P}}$ with this property one employs the
projection operator techniques to derive from the von Neumann
equation (\ref{NEUMANN}) for the total density matrix $\rho(t)$
exact and closed equations of motion for its projection
${\mathcal{P}}\rho(t)$ (see Sec.~\ref{Sec-Master}). The basic idea
underlying this approach is the following. With an appropriate
choice for the projection superoperator one intends to obtain a
description of the dynamics of system states which is much simpler
and much more efficient than the description by means of the full
density matrix $\rho(t)$. Thus, the map ${\mathcal{P}}$ expresses
the transition from the full representation in terms of the total
density matrix $\rho$ to a simplified, effective description
through a reduced set of dynamical variables defined by the
structure of the projection ${\mathcal{P}}\rho$.

Equation (\ref{PROJECTION}) is our first basic condition. It is
this condition which allows the formal application of the
projection operator techniques to open quantum systems. The
ultimate goal is to determine form the equations of motion for
${\mathcal{P}}\rho(t)$ the dynamics of the density matrix
$\rho_S(t)$ of the reduced open quantum system. To this end, we
need one further condition. Namely, whatever the form of the
projection superoperator is, we demand that ${\mathcal{P}}\rho$
contains the complete information required to reconstruct
$\rho_S$. We therefore impose the second basic condition:
\begin{equation} \label{CONSISTENCY}
 \rho_S \equiv {\mathrm{tr}}_E \rho = {\mathrm{tr}}_E
 \left\{ {\mathcal{P}} \rho \right\}.
\end{equation}
The first equation is just the definition of the reduced density
matrix which is obtained by taking the partial trace over the
environment. The second equation states that, in order to
determine $\rho_S$, we do not really need the full density matrix
of the total system, but only its projection ${\mathcal{P}} \rho$.
Hence, the reduced density matrix $\rho_S(t)$ is found by taking
the environmental trace of the equations of motion for
${\mathcal{P}}\rho(t)$.

Within the standard approach to the dynamics of open systems using
projection operator techniques one defines a projection
superoperator of the form:
\begin{equation} \label{PROJ-STANDARD}
 {\mathcal{P}} \rho = ({\mathrm{tr}}_E \rho) \otimes \rho_E,
\end{equation}
where $\rho_E$ is some fixed environmental density matrix referred
to as the reference state. This superoperator clearly satisfies
our basic conditions (\ref{PROJECTION}) and (\ref{CONSISTENCY}).
By use of this ${\mathcal{P}}$ the total state of the system is
represented by means of the tensor product state $\rho_S \otimes
\rho_E$. Regarding the reference state $\rho_E$ as fixed, one uses
the reduced density matrix $\rho_S(t)$ as the dynamical variable.
Applying the projection operator technique one then finds a master
equation for the reduced density matrix $\rho_S$ whose
coefficients are given by certain multitime correlation functions
defined by averages with respect to the reference state $\rho_E$.
In particular, the master equation obtained in second order of the
coupling yields the Born approximation of the dynamics which
involves certain two-time environmental correlation functions.

Our new class of correlated projection superoperators is obtained
as follows. We take any orthogonal decomposition of the unit
operator $I_E$ on the state space of the environment, i.~e., a
collection of projection operators $\Pi_a$ on ${\mathcal{H}}_E$
which satisfy
\begin{equation}
 \Pi_a \Pi_b = \delta_{ab} \Pi_b, \qquad \sum_a \Pi_a = I_E.
\end{equation}
Then we can define a linear map by means of
\begin{equation} \label{PROJ-NEW}
 {\mathcal{P}} \rho = \sum_a
 {\mathrm{tr}}_E \left\{ \Pi_a \rho \right\} \otimes \frac{1}{N_a}\Pi_a,
\end{equation}
where $N_a={\mathrm{tr}}_E \left\{ \Pi_a \right\}$. It is again
easy to verify that this superoperator fulfills the requirements
(\ref{PROJECTION}) and (\ref{CONSISTENCY}). By contrast to the
standard projection (\ref{PROJ-STANDARD}) which uses a
representation by a tensor product state, the new projection
(\ref{PROJ-NEW}) employs a set of un-normalized density matrices
\begin{equation}
 \rho^{(a)}_S = {\mathrm{tr}}_E \left\{ \Pi_a \rho \right\}
\end{equation}
in order to describe the states of the composite system. The set
of the density matrices $\rho^{(a)}_S(t)$ therefore represents the
dynamical variables defined by the projection superoperator
(\ref{PROJ-NEW}). Applying the projection operator technique one
is then led to a coupled system of equations of motion for the
$\rho^{(a)}_S(t)$, from which one obtains the reduced density
matrix itself by means of the relation (\ref{CONSISTENCY}):
\begin{equation} \label{RED-DENSITY}
 \rho_S(t)  = {\mathrm{tr}}_E \left\{ {\mathcal{P}}\rho(t) \right\}
 = \sum_a \rho^{(a)}_S(t).
\end{equation}

In the theory of entanglement (see, e.~g., Ref.~\cite{ALBER}) a
state of the form given by Eq.~(\ref{PROJ-NEW}) is called
separable or classically correlated \cite{WERNER}. The approach
based on a projection of this form thus tries to approximate the
total system's state through a classically correlated but
non-factorizing state. Of course, one can also construct
projection superoperators which lead to non-separable (entangled)
states ${\mathcal{P}}\rho$. The examples discussed below belong to
the classes of projection operators defined by
Eqs.~(\ref{PROJ-STANDARD}) and (\ref{PROJ-NEW}).

\subsection{Equations of motion}\label{Sec-Master}
Basically, there are two variants of the projection operator
technique. The first one is the prominent Nakajima-Zwanzig method
\cite{NAKAJIMA,ZWANZIG}. It leads to a first-order
integro-differential equation for ${\mathcal{P}}\rho(t)$ which
contains a time integration over the past system history involving
a certain memory kernel. The second variant is known as
time-convolutionless (TCL) projection operator technique
\cite{SHIBATA1,SHIBATA2} which yields a time-local equation of
motion for ${\mathcal{P}}\rho(t)$. We shall use this second
variant of the projection operator technique in the present paper.
Its advantage is that in any order of the coupling one only has to
solve a first-order differential equation which is local in time.
It should be emphasized, however, that the general considerations
developed here may also be applied to the Nakajima-Zwanzig
projection operator technique.

The TCL projection operator method leads to an equation of motion
for the projection ${\mathcal{P}}\rho(t)$ which is of the general
form:
\begin{equation} \label{TCL-MASTER}
 \frac{d}{dt} {\mathcal{P}}\rho(t) = {\mathcal{K}}(t)
 {\mathcal{P}} \rho(t) + {\mathcal{I}}(t) {\mathcal{Q}} \rho(0).
\end{equation}
This is an exact inhomogeneous linear differential equation of
first order. Both the TCL generator ${\mathcal{K}}(t)$ of the
linear part and the inhomogeneity ${\mathcal{I}}(t)$ are
explicitly time-dependent superoperators.

The inhomogeneous part of Eq.~(\ref{TCL-MASTER}) is determined by
the projection ${\mathcal{Q}}\rho(0)$ of the initial state, where
${\mathcal{Q}}=I-{\mathcal{P}}$ is the projection superoperator
complementary to ${\mathcal{P}}$, and $I$ denotes the unit map. We
observe that the inhomogeneous term vanishes if the initial state
satisfies the relation ${\mathcal{Q}}\rho(0)=0$, i.~e., if
\begin{equation} \label{INIT-STATE}
 {\mathcal{P}}\rho(0) = \rho(0).
\end{equation}
Obviously, this relation simplifies the equation of motion
(\ref{TCL-MASTER}). In the case of the standard projection
(\ref{PROJ-STANDARD}) it implies that $\rho(0)$ is an
un-correlated tensor product state, i.~e.
$\rho(0)=\rho_S(0)\otimes\rho_E$. In the case of the projection
(\ref{PROJ-NEW}), however, Eq.~(\ref{INIT-STATE}) merely implies
that the initial state is of the correlated form given by the
structure of the projection superoperator (\ref{PROJ-NEW}).

In general, the TCL generator ${\mathcal{K}}(t)$ and the
inhomogeneity ${\mathcal{I}}(t)$ are extremely complicated
objects, and the exact solution of Eq.~(\ref{TCL-MASTER}) is as
difficult as the solution of the full von Neumann equation for the
total system. However, Eq.~(\ref{TCL-MASTER}) can be used as a
starting point of a systematic perturbation expansion with respect
to the strength of the interaction Hamiltonian $V$. With the help
of the TCL technique one derives a closed expression for the
corresponding expansion of the TCL generator:
\begin{equation}
 {\mathcal{K}}(t) = \sum_{n=1}^{\infty} {\mathcal{K}}_n(t).
\end{equation}
The $n$-th-order contribution is given by
\begin{eqnarray}
 {\mathcal{K}}_n(t) &=& \int_0^t dt_1  \int_0^{t_1} dt_2 \ldots
 \int_0^{t_{n-2}} dt_{n-1} \nonumber \\
 &~& \times \langle {\mathcal{L}}(t) {\mathcal{L}}(t_1) {\mathcal{L}}(t_2)
 \ldots {\mathcal{L}}(t_{n-1}) \rangle_{\mathrm{oc}}.
\end{eqnarray}
The quantities
\begin{eqnarray*}
 \lefteqn{ \langle {\mathcal{L}}(t) {\mathcal{L}}(t_1) {\mathcal{L}}(t_2) \ldots
 {\mathcal{L}}(t_{n-1}) \rangle_{\mathrm{oc}} \equiv } \\
 && \sum (-1)^q \mathcal{P} {\mathcal{L}}(t) \ldots {\mathcal{L}}(t_i)
 \mathcal{P} {\mathcal{L}}(t_j) \ldots {\mathcal{L}}(t_k)  \mathcal{P}
 \ldots {\mathcal{P}}
\end{eqnarray*}
are known as ordered cumulants
\cite{KUBO63,ROYER72,KAMPEN1,KAMPEN2} and are defined by the
following rules: (1) Write a string of the form
${\mathcal{P}}{\mathcal{L}}\ldots{\mathcal{L}}{\mathcal{P}}$ with
$n$ factors of ${\mathcal{L}}$ in between two ${\mathcal{P}}$'s.
(2) Insert an arbitrary number $q$ of factors ${\mathcal{P}}$
between the ${\mathcal{L}}$'s such that at least one
${\mathcal{L}}$ stands between two successive ${\mathcal{P}}$'s.
The resulting expression is multiplied by a factor $(-1)^q$ and
all ${\mathcal{L}}$'s are furnished with a time argument: The
first time argument is always $t$. The remaining ${\mathcal{L}}$'s
carry any permutation of the time arguments
$t_1,t_2,\ldots,t_{n-1}$ with the only restriction that the time
arguments in between two successive ${\mathcal{P}}$'s must be
ordered chronologically. In the above expression we thus have $t
\geq \ldots \geq t_i$, $t_j \geq \ldots \geq t_k$, etc. (3)
Finally, the ordered cumulant is obtained by a summation over all
possible insertions of ${\mathcal{P}}$ factors and over all
allowed distributions of the time arguments.

In many physical applications it may be assumed that the relations
\begin{equation} \label{ODD}
 {\mathcal{P}} {\mathcal{L}}(t) {\mathcal{L}}(t_1) \ldots
 {\mathcal{L}}(t_{2n}) {\mathcal{P}}  = 0
\end{equation}
hold, which means that any string containing an odd number of
${\mathcal{L}}$'s between successive factors of ${\mathcal{P}}$
vanishes. Following the above rules one then finds that all
odd-order contributions ${\mathcal{K}}_{2n+1}(t)$ vanish, while
the second and the fourth-order terms take the form:
\begin{equation} \label{K2}
 {\mathcal{K}}_2(t) = \int_0^{t} dt_1
 {\mathcal{P}} {\mathcal{L}}(t) {\mathcal{L}}(t_1) {\mathcal{P}},
\end{equation}
and
\begin{eqnarray} \label{K4}
 {\mathcal{K}}_4(t) &=&
 \int_0^tdt_1\int_0^{t_1}dt_2\int_0^{t_2}dt_3 \nonumber \\
 &~& \times \Big(
 {\mathcal P}{\mathcal{L}}(t){\mathcal{L}}(t_1){\mathcal{L}}(t_2)
 {\mathcal{L}}(t_3){\mathcal P} \nonumber \\
 &~& \;\;\;
 -{\mathcal P}{\mathcal{L}}(t){\mathcal{L}}(t_1){\mathcal P}
  {\mathcal{L}}(t_2){\mathcal{L}}(t_3){\mathcal P} \nonumber \\
 &~& \;\;\;
 -{\mathcal P}{\mathcal{L}}(t){\mathcal{L}}(t_2){\mathcal P}
  {\mathcal{L}}(t_1){\mathcal{L}}(t_3){\mathcal P} \nonumber \\
 &~& \;\;\;
 -{\mathcal P}{\mathcal{L}}(t){\mathcal{L}}(t_3){\mathcal P}
  {\mathcal{L}}(t_1){\mathcal{L}}(t_2){\mathcal P}
 \Big).
\end{eqnarray}

The performance of the formal expansion outlined above strongly
depends, of course, on the choice of the projection superoperator
${\mathcal{P}}$. In other words, the quality of the approximation
obtained by truncating the expansion at a given order $n$
crucially depends on the structure of the chosen projection. It is
important to note that the technique yields an expansion of a
certain system of equations of motion, and not of the reduced
system's density matrix itself. Taking different projection
superoperators one uses entirely different sets of dynamical
variables which obey completely different equations of motion.
 Hence, changing the projection superoperator amounts to changing
the set of dynamical variables and the whole structure of the
equations of motion, and to a non-perturbative re-organization of
the expansion. It may even happen that the solution of the
equations of motion in a given order for one particular projection
represents the solution to all orders for another projection. This
point will be illustrated in Sec.~\ref{Sec-Appl} by means of a
specific system-reservoir model.

\section{Hilbert space averaging approach to the reduced
         dynamics}\label{HAM-RED}

\subsection{The Hilbert space average method}\label{ham}

The Hilbert space average method (HAM) is in essence a technique
to produce guesses for the values of quantities defined as
functions of a wavefunction $\ket{\psi}$ if $\ket{\psi}$ itself is
not known in full detail, only some features of it. In particular
it produces a guess for the expectation value
$\bra{\psi}\hat{A}\ket{\psi}$ if the only information about
$\ket{\psi}$ is a set of different expectation values
$\bra{\psi}\hat{B}_n\ket{\psi}=B_n$. Such a statement naturally
has to be a guess since there are in general many different
$\ket{\psi}$ that are in accord with the given set of $B_n$ but
produce possibly different values for
$\bra{\psi}\hat{A}\ket{\psi}$. The question now is whether the
distribution of $\bra{\psi}\hat{A}\ket{\psi}$'s produced by the
respective set of $\ket{\psi}$'s is broad or whether almost all
those $\ket{\psi}$'s yield $\bra{\psi}\hat{A}\ket{\psi}$'s that
are approximately equal. It turns out that if the spectral width
of $\hat{A}$ is not too large and  $\hat{A}$ is high-dimensional
almost all individual $\ket{\psi}$ yield an expectation value
close to the mean of the distribution of
$\bra{\psi}\hat{A}\ket{\psi}$'s. In spite of this being crucial
for the following we refer the reader to \cite{GEMMER} for
details. To find that mean one has to average with respect to the
$\ket{\psi}$'s. We call this a Hilbert space average $A$ and
denote it as
\begin{equation} \label{eq:29}
 A=\Haver{\bra{\psi}\hat{A}\ket{\psi}}_{\{\bra{\psi}\hat{B}_n\ket{\psi}=B_n\}}.
\end{equation}
This expression stands for the average of
$\bra{\psi}\hat{A}\ket{\psi}$ over all $\ket{\psi}$ that feature
$\bra{\psi}\hat{B}_n\ket{\psi}=B_n$ but are uniformly distributed
otherwise. Uniformly distributed means invariant with respect to all
unitary transformations that leave $\bra{\psi}\hat{B}_n\ket{\psi}=B_n$
unchanged.
One may rewrite (\ref{eq:29}) as
\begin{equation} \label{eq:30}
 A = {\mathrm{tr}}\{\hat{A}\hat{\alpha}\}
 \quad \mbox{with} \quad
 \hat{\alpha} \equiv
 \Haver{\ket{\psi}\bra{\psi}}_{\{\bra{\psi}\hat{B}_n\ket{\psi}=B_n\}}.
\end{equation}
How is $\hat{\alpha}$ to be computed? Any unitary transformation
that leaves $\bra{\psi}\hat{B}_n\ket{\psi}=B_n$ invariant has to
leave $\hat{\alpha}$ invariant, i.~e.:
\begin{equation} \label{eq:31}
 e^{i\hat{G}}\hat{\alpha}e^{-i\hat{G}}=\hat{\alpha} \quad
 \mbox{with} \quad [\hat{G},\hat{B}_n]=0.
\end{equation}
This, however, can only be fulfilled if
$[\hat{G},\hat{\alpha}]=0$ and this leads to the general
form
\begin{equation} \label{eq:31a}
 \hat{\alpha}=\sum_n b_n \hat{B}_n.
\end{equation}
(In principal there could be addends of the form, e.~g.,
$\hat{B}_n\hat{B}_m$, etc., but since all $\hat{B}_n$ we are going to
consider below together
with zero form a group, those addends are already contained in the
above sum.)

Furthermore one of course has the following conditions:
\begin{equation} \label{eq:32}
 {\mathrm{tr}} \{ {\hat{\alpha}}\hat{B}_m \} = B_m.
\end{equation}
One thus obtains
\begin{equation} \label{eq:33}
 B_m = \sum_n {\mathrm{tr}} \{ \hat{B}_m \hat{B}_n \} b_n
\end{equation}
from which the $b_n$ may be determined. Thus, the construction of
a given Hilbert space average is defined with the help of
Eqs.~(\ref{eq:33}), (\ref{eq:31a}), and (\ref{eq:30}). According
to this scheme ``best guesses'' for certain expectation values
will be produced below. The explanation of HAM in full detail is
beyond the scope of this text and can be found in
\cite{Gemmer2003,GEMMER}.

\subsection{HAM, projection operators and dynamics}
In the following we explain how HAM can be used to produce the
reduced dynamics of a quantum system coupled to some environment,
just like the techniques described in Sect.~\ref{Sec-Projection}.
Consider the full system's pure state at some time $t$,
$\ket{\psi(t)}$. Let $\hat{D}(\tau)$ be a time evolution operator
describing the evolution of the system for a short time, i.~e.,
$\ket{\psi(t+\tau)}=\hat{D}(\tau)\ket{\psi(t)}$. This allows for
the computation of a set of observables $\hat{B}_n$ at time
$t+\tau$:
\begin{equation} \label{eq:34}
 B_n(t+\tau) =
 \bra{\psi(t)}\hat{D}^{\dagger}(\tau)\hat{B}_n\hat{D}(\tau)\ket{\psi(t)}.
\end{equation}
Now assume that rather than $\ket{\psi(t)}$ itself only the set of
expectation values $B_n(t)=\bra{\psi(t)}\hat{B}_n\ket{\psi(t)}$ is
known. The application of HAM produces a guess for the
$B_n(t+\tau)$ based on the  $B_n(t)$:
\begin{equation} \label{eq:35}
 B_n(t+\tau)\approx
 \Haver{\bra{\phi}\hat{D}^{\dagger}(\tau)\hat{B}_n\hat{D}(\tau)
 \ket{\phi}}_{\{\bra{\phi}\hat{B}_n\ket{\phi}=B_n(t)\}}.
\end{equation}
(Note that here the $\ket{\phi}$ appear rather than the
$\ket{\psi(t)}$ because those are not actually realized states but
denote the set of states over which the Hilbert space average has
to be taken.) Iterating this scheme, i.~e., taking the
$B_n(t+\tau)$ for the $B_n(t)$ of the next step allows for the
stepwise computation of the evolution of the $B_n$'s. If the set
of the $B_n$ is chosen such that it determines the local state of
the considered quantum system completely this technique produces
the local reduced dynamics. The result is of course just like HAM
itself only a best guess, but for appropriate systems this guess
can be rather accurate.

Here, the $\hat{B}_n$'s are chosen specifically as operators
corresponding to elements of the reduced density matrix of the
considered system and the occupation probability of ``energy
bands'' of the environment:
\begin{equation} \label{eq:36}
 \hat{B}_n \equiv \hat{B}_{ija} \equiv \ket{i}\bra{j} \otimes
 \Pi_a,
\end{equation}
where $\ket{i},\ket{j}$ are energy eigenstates of the considered
system and $ \Pi_a$ is as described in
Sect.~\ref{Sec-Superoperators} a projector, projecting out the
energy eigenstates of the environment belonging to an interval
$\Delta E_a$ around some mean band energy $E_a$ labelled by the
index $a$. Let $\bra{\psi}\hat{B}_{ija}\ket{\psi} \equiv B_{ija}$,
then one gets for the elements $\rho_{ij}$ of the reduced density
matrix:
\begin{equation}
 \label{eq:37}
\rho_{ij}=\sum_a B_{jia}.
\end{equation}
Thus, the above defined set of expectation values determines the
reduced state of the system completely. The set of states
$\hat{\alpha}$ which belong to the Hilbert space average defined
by the $B_{jia}$ [in the sense of Eq.~(\ref{eq:30})] is, following
the scheme described in Sect.~\ref{ham}, found to be
\begin{equation}
 \label{eq:38}
\hat{\alpha}=\sum_{ija}\frac{B_{jia}}{N_a}\hat{B}_{ija}.
\end{equation}
(This turns out to be the same state one would have gotten from
minimizing the purity under the subsidiary condition set by the
given expectation values $B_{jia}$.) A comparison with the
considerations of Sec.~\ref{Sec-Superoperators} reveals that
$\hat{\alpha}$ has exactly the form produced by the application of
the projection superoperator $\mathcal{P}$ [see
Eq.~(\ref{PROJ-NEW})]. Exploiting Eqs.~(\ref{eq:29}),
(\ref{eq:30}), and (\ref{eq:38}) one can write a specific form of
Eq.~(\ref{eq:35}) for this case:
\begin{eqnarray} \label{eq:39}
  \lefteqn{ B_{ija}(t+\tau) \approx } \\
 && \sum_{lmb}\frac{1}{N_b}{\mathrm{tr}}\left\{\ket{m}\bra{l}\Pi_b\hat{D}^{\dagger}(\tau)
 \ket{i}\bra{j}\Pi_a\hat{D}(\tau)\right\}B_{lmb}(t).\nonumber
\end{eqnarray}
Working in the interaction picture the dynamics of the full system
is controlled by the interaction $V(t)$. The time evolution is
generated by the corresponding Dyson series. Thus, assuming weak
interactions Eq.~(\ref{eq:39}) may be evaluated to second order in
the interaction strength using an appropriately truncated Dyson
series for $\hat{D}(\tau)$. This yields after extensive but rather
straightforward calculations for the expectation values
corresponding to diagonal elements:
\begin{eqnarray} \label{eq:40}
 \lefteqn{ B_{iia}(t+\tau) = } \\
 && B_{iia}(t)+\sum_{jb}f(ijab,\tau)\left(\frac{B_{jjb}(t)}{N_b}-\frac{B_{iia}(t)}{N_a}\right),\nonumber
\end{eqnarray}
and for the expectation values corresponding to off-diagonal
elements:
\begin{eqnarray} \label{eq:41}
 \lefteqn{ B_{ija}(t+\tau) = } \\
 && B_{ija}(t)-\frac{1}{2}\frac{B_{ija}(t)}{N_a}\sum_{kb}(f(ikab,\tau)
 +f(kjab,\tau)), \nonumber
\end{eqnarray}
where
\begin{eqnarray} \label{eq:42}
 f(ijab,\tau) &\equiv& 2\int_0^{\tau} d\tau' \int_0^{\tau'} d\tau'' \\
 && \times {\mathrm{tr}}
 \left\{ \Pi_a\bra{i}V(\tau'')\ket{j}\Pi_b\bra{j}V(0)\ket{i} \right\}.
 \nonumber
\end{eqnarray}
Those $f$'s are essentially integrals over the same environmental
temporal correlation functions that appear in the memory kernels
of standard projection operator techniques. But here they
explicitly correspond to transitions between different energy
subspaces of the environment. (In Eqs.~(\ref{eq:40}) and
(\ref{eq:41}) we assumed that correlation functions vanish unless
they refer to correlations between parts of the interaction that
are adjoints of each other as in Eq.~(\ref{eq:42}). We furthermore
assumed ${\mathrm{tr}}\{\Pi_a\bra{i}V(\tau)\ket{j}\}=0$. Both
conditions are not necessarily fulfilled but apply to the concrete
model analyzed below.) Those correlation functions typically
feature (short) decay times, i.~e, integrating them twice yields
functions which increase linear in time after the corresponding
decay time. Thus, for $\tau$ larger than the decay time
Eq.~(\ref{eq:42}) may be written as
\begin{equation} \label{eq:43}
 f(ijab,\tau)\approx N_b\gamma(ijab)\tau,
\end{equation}
where $\gamma(ijab)$ has to be computed from Eq.~(\ref{eq:42}) but
typically corresponds to a transition rate as obtained from
Fermi's Golden Rule. Especially it will only be non-zero for
$E_i-E_j\approx E_a-E_b$ for otherwise the correlation function
oscillates rapidly before it decays and hence the corresponding
integrals vanish.

Inserting Eq.~(\ref{eq:43}) into Eqs.~(\ref{eq:40}) and
(\ref{eq:41}) and assuming that the decay times of the correlation
functions are small compared to the resulting decay times of the
system (which are of the order of $1/\gamma(ijab)$), one can
transform the iteration scheme into a set of differential
equations:
\begin{align} \label{eq:44}
 \frac{d}{dt}B_{iia}& = \sum_{jb} \gamma(ijab)
 \left( B_{jjb}-\frac{N_b}{N_a}B_{iia} \right), \\
 \frac{d}{dt}B_{ija}&=-\frac{1}{2}B_{ija}\sum_{kb}
 \left( \gamma(kiba)+\gamma(jkba) \right).
\end{align}
This set of differential equations obviously determines the
reduced dynamics of the considered system. It produces an
exponential decay to an equilibrium state. Again those dynamics
are only a guess, but as a guess they are valid for any initial
state regardless of whether it is pure, correlated, entangled,
etc. In contrast to the standard Nakajima-Zwanzig and TCL methods
where initial states generally produce an inhomogeneity [see
Eq.~(\ref{TCL-MASTER})] which may be difficult to handle, HAM
allows for a direct guess on the typical behavior of the system.
However, a crucial condition for the application of the above
scheme is that the decay times of the correlations are short
enough such that even for larger times the evolution is well
described by a Dyson series truncated at second order. This means
that the scheme will break down altogether if the interaction is
too strong.

\section{Application}\label{Sec-Appl}

\subsection{The model}

\begin{figure}[htb]
\includegraphics[width=0.8\linewidth]{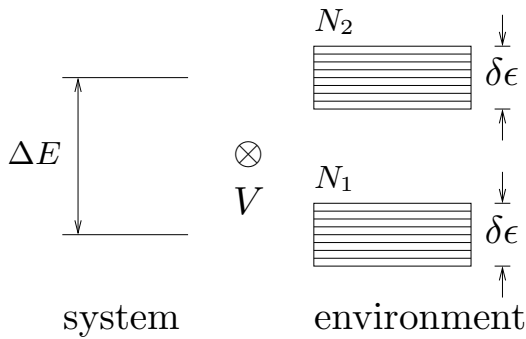}
\caption{a two-state system coupled to an environment consisting
of two energy bands with a finite number of
levels.\label{ModelFig}}
\end{figure}

To illustrate the general considerations of the previous sections
we investigate the model of a two-state system $S$ which is
coupled to an environment $E$ \cite{GEMMER}. The environment
consists of a large number of energy levels arranged in two energy
bands of the same width $\delta\varepsilon$. The levels of each
band are equidistant. The lower energy band contains $N_1$ levels,
the upper band $N_2$ levels. The transition of the two-state
system is in resonance with the distance $\Delta E$ between the
bands (see Fig.~\ref{ModelFig}).

The total Schr\"odinger picture Hamiltonian of the model is taken
to be $H = H_0 + V$, where
\begin{eqnarray} \label{MODEL-H0}
 H_0 &=& \Delta E \sigma_z + \sum_{n_1} \frac{\delta\varepsilon}{N_1}
 n_1 |n_1\rangle\langle n_1| \nonumber \\
 &~& + \sum_{n_2}
 \left( \Delta E + \frac{\delta\varepsilon}{N_2}n_2 \right)
 |n_2\rangle\langle n_2|,
\end{eqnarray}
and
\begin{equation} \label{MODEL-V}
 V = \lambda \sum_{n_1,n_2} c(n_1,n_2) \sigma_+
 |n_1\rangle\langle n_2| + {\mbox{h.c.}}
\end{equation}
Here and in the following the index $n_1$ labels the levels of the
lower energy band and $n_2$ the levels of the upper band.
$\sigma_z$ and $\sigma_+$ are standard Pauli matrices. The overall
strength of the interaction is parameterized by the constant
$\lambda$. The coupling constants $c(n_1,n_2)$ are independent and
identically distributed complex Gaussian random variables
satisfying:
\begin{eqnarray}
 \langle c(n_1,n_2) \rangle &=& 0, \label{AV1} \\
 \langle c(n_1,n_2) c(n'_1,n'_2) \rangle &=& 0, \label{AV2} \\
 \langle c(n_1,n_2) c^*(n'_1,n'_2) \rangle &=& \delta_{n_1,n'_1}
 \delta_{n_2,n'_2}. \label{AV3}
\end{eqnarray}
Transforming to the interaction picture we get the von Neumann
equation (\ref{NEUMANN}) with the interaction picture Hamiltonian:
\begin{equation}
 V(t) = \sigma_+ B(t) + \sigma_- B^{\dagger}(t),
\end{equation}
where
\begin{equation}
 B(t) = \lambda \sum_{n_1,n_2} c(n_1,n_2) e^{-i\omega(n_1,n_2)t}
 |n_1\rangle\langle n_2|
\end{equation}
and
\begin{equation}
 \omega(n_1,n_2) = \delta\varepsilon\left(\frac{n_2}{N_2}-\frac{n_1}{N_1} \right).
\end{equation}

\subsection{The standard approach}

\subsubsection{Projection superoperator}
In the standard approach one uses a projection superoperator of
the form given by Eq.~(\ref{PROJ-STANDARD}). Let us denote the
projection onto the lower (upper) band by $\Pi_1$ ($\Pi_2$):
\begin{eqnarray}
 \Pi_1 &=& \sum_{n_1} |n_1\rangle\langle n_1|, \\
 \Pi_2 &=& \sum_{n_2} |n_2\rangle\langle n_2|.
\end{eqnarray}
We consider initial states for which only the lower band is
occupied: $\rho(0)=\rho_S(0)\otimes\Pi_1/N_1$. Hence, if we take
the reference state
\begin{equation}
 \rho_E = \frac{1}{N_1} \Pi_1,
\end{equation}
we have
\begin{equation}
 {\mathcal{P}} \rho = ({\mathrm{tr}}_E \rho) \otimes \rho_E = \rho_S
 \otimes \frac{1}{N_1} \Pi_1,
\end{equation}
and
\begin{equation} \label{INIT-COND}
 {\mathcal{P}}\rho(0) = \rho(0).
\end{equation}
In the following we write the elements of the reduced density
matrix as
\begin{equation}
 \rho_{ij}(t) = \langle i |\rho_S(t)|j\rangle, \qquad i,j=0,1,
\end{equation}
where $|0\rangle$ and $|1\rangle$ denote the lower and the upper
level of the two-state system, respectively. If follows from
Eq.~(\ref{INIT-COND}) that the inhomogeneous term of the TCL
master equation (\ref{TCL-MASTER}) vanishes. It can also be
verified easily with the help of the above forms for the
projection superoperator and the interaction Hamiltonian that the
condition (\ref{ODD}) holds true. Thus, the TCL generator
${\mathcal{K}}(t)$ contains only the contributions from even
orders of the coupling strength $\lambda$.

\subsubsection{TCL master equation of second order}
From the expression (\ref{K2}) for the second-order contribution
of the TCL generator we find:
\begin{eqnarray*}
 {\mathcal{K}}_2(t) {\mathcal{P}} \rho(t) &=&
 \int_0^t dt_1 f_2(t,t_1) \nonumber \\
 &~& \times \left[ 2\sigma_- \rho_S(t) \sigma_+
 - \{ \sigma_+\sigma_-, \rho_S(t) \}\right] \otimes
 \rho_E,
\end{eqnarray*}
where $\{\cdot,\cdot\}$ denotes the anticommutator and
\begin{eqnarray} \label{2-POINT}
 f_2(t,t_1) &=& \left\langle {\mathrm{tr}}_E
 \{ B(t)B^{\dagger}(t_1) \rho_E \} \right\rangle \nonumber \\
 &\equiv& \gamma_2 h(t-t_1)
\end{eqnarray}
is the two-point environmental correlation function with
\begin{equation} \label{DEF-GAMMA2}
 \gamma_2 = \frac{2\pi\lambda^2N_2}{\delta\varepsilon}.
\end{equation}
The angular brackets in Eq.~(\ref{2-POINT}) denote the average
over the random couplings $c(n_1,n_2)$ which is determined by use
of the relations (\ref{AV1})-(\ref{AV3}). The function $h(\tau)$
introduced in Eq.~(\ref{2-POINT}) is then found to be
\begin{equation} \label{h-def}
 h(\tau) = \frac{\delta\varepsilon}{2\pi}
 \frac{\sin^2(\delta\varepsilon\cdot\tau/2)}{(\delta\varepsilon\cdot\tau/2)^2},
\end{equation}
where we have assumed a constant finite density of states for the
environmental energy bands. This function exhibits a sharp peak of
width $\delta\varepsilon^{-1}$ at $\tau=0$ and may be approximated
by a delta function for times $t$ which are large compared to the
inverse band width, i.~e., for $\delta\varepsilon \cdot t \gg 1$
we may approximate:
\begin{equation}
 f_2(t,t_1) \approx \gamma_2 \delta(t-t_1).
\end{equation}
This yields the second-order master equation for the reduced
density matrix:
\begin{equation} \label{TCL2}
 \frac{d}{dt} \rho_S(t)
 = \gamma_2 \left[ \sigma_- \rho_S(t) \sigma_+
 - \frac{1}{2} \{ \sigma_+\sigma_-, \rho_S(t) \}\right].
\end{equation}
This is a quantum Markovian master equation in Lindblad form,
where the quantity $\gamma_2$ represents the Markovian relaxation
rate.

On the ground of the second order approximation one could naively
expect that the master equation (\ref{TCL2}) provides a reasonable
approximation of the reduced system's dynamics if the relaxation
rate $\gamma_2$ is small compared to the band width:
\begin{equation} \label{MARKOV}
 \gamma_2 \ll \delta \varepsilon.
\end{equation}
However, we are going to demonstrate that this is {\textit{not}}
true. A comparison with numerical simulations of the full
Schr\"odinger equation and with the prediction of HAM shows that
the long time dynamics is not correctly reproduced by this master
equation.

\begin{figure}[htb]
\includegraphics[width=0.95\linewidth]{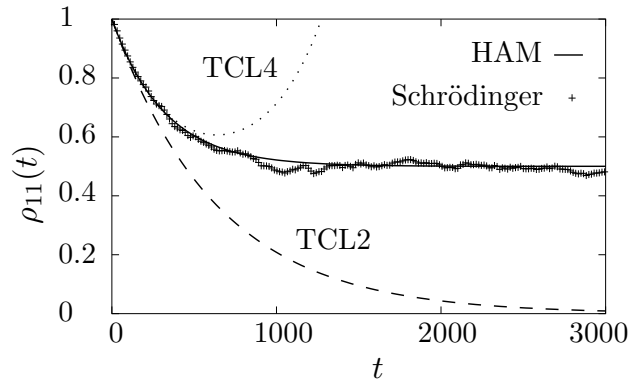}
\caption{comparison of the numerical solution of the Schr\"odinger
equation with the approximations given by HAM [Eq.~(\ref{HAM2})]
and by the second and the fourth order of the standard TCL
expansion [Eqs.~(\ref{POPTCL2}) and (\ref{POPTCL4})]. Parameters:
$N_1=N_2=500$, $\delta\varepsilon=0.5$, and $\lambda=5\cdot
10^{-4}$.\label{TCLHAMfig}}
\end{figure}

The approximation of HAM leads to the following expression for the
population of the upper level:
\begin{equation} \label{HAM2}
 \rho_{11}(t) = \rho_{11}(0) \left[ \frac{\gamma_1}{\gamma_1+\gamma_2}
 + \frac{\gamma_2}{\gamma_1+\gamma_2}
 e^{-(\gamma_1+\gamma_2)t} \right],
\end{equation}
where $\gamma_2$ is defined by Eq.~(\ref{DEF-GAMMA2}) and we have
introduced a further relaxation rate:
\begin{equation} \label{DEF-GAMMA1}
 \gamma_1 = \frac{2\pi\lambda^2N_1}{\delta\varepsilon}.
\end{equation}
On the other hand, the TCL master equation (\ref{TCL2}) gives:
\begin{equation} \label{POPTCL2}
 \rho_{11}(t) = \rho_{11}(0) e^{-\gamma_2t}.
\end{equation}
Thus, the TCL master equation predicts an exponential relaxation
of the populations to zero, while the solution obtained by means
of HAM approaches the stationary population
\begin{equation}
 \rho_{11}^{\mathrm{stat}} = \rho_{11}(0) \frac{\gamma_1}{\gamma_1+\gamma_2}.
\end{equation}

To judge the quality of the various approximations we have
performed numerical solutions of the full Schr\"odinger equation
corresponding to the Hamiltonian defined by Eqs.~(\ref{MODEL-H0})
and (\ref{MODEL-V}). The initial state has been taken to be of the
form $|1\rangle \otimes |\chi\rangle$, where the environmental
state $|\chi\rangle$ represents a superposition of the states
$|n_1\rangle$ of the lower band with independent Gaussian
distributed random amplitudes of zero mean and equal variances.
For certain parameter ranges we find an excellent agreement of the
HAM prediction with the simulation results. An example is shown in
Fig.~\ref{TCLHAMfig}. Note that for the parameters of this figure
we have $\gamma_2/\delta\varepsilon = 3\cdot 10^{-3}$, such that
the standard Markov condition (\ref{MARKOV}) is very well
satisfied.

We conclude that although the standard Markov condition
(\ref{MARKOV}) is fulfilled the Markovian master equation
(\ref{TCL2}) does not yield a good approximation of the dynamics
for intermediate and long times. In particular, its prediction for
the stationary state is totally wrong. The important point to note
is that, in order to judge the quality of the Markov
approximation, an analysis of the contributions of higher orders
is indispensable. We also note that the same problem occurs if one
uses the Nakajima-Zwanzig master equation.

\subsubsection{The master equation of fourth order
               and failure of the Born-Markov approximation}
To understand the failure of the Born-Markov approximation we
investigate the fourth order of the TCL expansion. The
contribution of fourth order to the TCL generator is given by
Eq.~(\ref{K4}). One finds that this contribution is determined by
the two-point correlation function (\ref{2-POINT}) and by the
four-point correlation function:
\begin{equation}
 f_4(t,t_1,t_2,t_3) = \left\langle {\mathrm{tr}}_E
 \{ B(t)B^{\dagger}(t_1) B(t_2)B^{\dagger}(t_3) \rho_E \}
 \right\rangle.
\end{equation}
The analysis shows that this function has two sharp peaks of width
$\delta\varepsilon^{-1}$ at $t=t_1$, $t_2=t_3$ and at $t_1=t_2$,
$t=t_3$, and may be approximated, under the conditions of the
previous section, by the expression:
\begin{eqnarray} \label{2-PEAKS}
 f_4(t,t_1,t_2,t_3) &\approx&
 \gamma^2_2 \delta(t-t_1) \delta(t_2-t_3) \nonumber \\
 &~& + \gamma_1\gamma_2 \delta(t-t_3) \delta(t_1-t_2),
\end{eqnarray}
where $\gamma_{1,2}$ are defined by Eqs.~(\ref{DEF-GAMMA1}) and
(\ref{DEF-GAMMA2}).

The double-peak structure of the four-point correlation expressed
by Eq.~(\ref{2-PEAKS}) has decisive consequences. With the help of
the correlation functions given above the master equation of
fourth order in the coupling is found to be:
\begin{eqnarray} \label{TCL4}
 \frac{d}{dt} \rho_S(t)
 &=& \Gamma(t) \left[ \sigma_- \rho_S \sigma_+
 - \frac{1}{2} \{ \sigma_+\sigma_-, \rho_S \}\right] \\
 &+& \tilde{\Gamma}(t) \left[ \sigma_+\sigma_- \rho_S
 \sigma_+ \sigma_-
 - \frac{1}{2} \{ \sigma_+\sigma_-, \rho_S \}\right], \nonumber
\end{eqnarray}
where
\begin{equation}
 \Gamma(t) = \gamma_2 (1-\gamma_1 t), \qquad
 \tilde{\Gamma}(t) = \gamma_1\gamma_2 t.
\end{equation}
Equation (\ref{TCL4}) is a master equation with time-dependent
relaxation rates $\Gamma(t)$ and $\tilde{\Gamma}(t)$. The
second-order contribution to the rate $\Gamma(t)$ is given by
$\gamma_2$, while the fourth order yields the contribution
$\gamma_2 \cdot (\gamma_1t)$. Therefore, the fourth-order term of
the TCL generator is small compared to the second-order term only
if the times $t$ considered satisfy the additional condition
\begin{equation}
 \gamma_1 t \ll 1.
\end{equation}
Thus we see that the occurrence of terms proportional to $t$ is
responsible for strong deviations from the Markovian behavior.
These terms are due to the double-peak structure of the four-point
correlation function $f_4$. We remark that this structure is
markedly different to the usual situation of the coupling of an
open system to a Bosonic field vacuum, for example. In this case
$f_4$ has only a single peak and, hence, the above phenomenon of
strong deviations from the Markovian dynamics for weak couplings
does not occur.

The master equation (\ref{TCL4}) yields the coherences:
\begin{equation}
 \rho_{01}(t) = \rho_{01}(0) e^{-\gamma_2t/2},
\end{equation}
and the populations:
\begin{equation} \label{POPTCL4}
 \rho_{11}(t) = \rho_{11}(0) e^{-\gamma_2t +
 \gamma_1\gamma_2t^2/2}.
\end{equation}
For $\gamma_{1,2}t \ll 1$ we find the expansion:
\[
 \rho_{11}(t) = \rho_{11}(0) \left[ 1 - \gamma_2t
 + \frac{1}{2}\gamma_2(\gamma_1+\gamma_2)t^2 + \ldots \right],
\]
which is seen to coincide with the corresponding short-time
expansion of the HAM approximation given by Eq.~(\ref{HAM2}). Thus
we conclude that the TCL expansion based on the standard
projection reproduces the short-time behavior predicted by HAM
within the given orders. Correspondingly, the TCL approximation of
fourth order clearly improves the approximation for short times,
but leads to un-physical results for longer times and diverges in
the limit $t\rightarrow\infty$, as is illustrated in
Fig.~\ref{TCLHAMfig}. We note that a similar situation occurs for
the spin star model studied in Ref.~\cite{BURGARTH}.

Summarizing, the fourth order clearly indicates that the TCL
expansion does not converge uniformly in $t$. It only provides a
short-time expansion of the dynamics. As a result of the emergence
of terms which are given by powers of $t$, it is impossible to
obtain valid predictions on the long-time dynamics if one
truncates the TCL series at any finite order.

\subsection{TCL expansion using the correlated projection
            superoperator}\label{NEWTCL}

In view of the above analysis the following question arises: Is it
possible to construct a new projection superoperator
${\mathcal{P}}$ whose corresponding TCL expansion yields the full
prediction of HAM already in lowest order, and leads to a
systematic expansion around HAM in higher orders? To answer this
question we consider the following projection superoperator:
\begin{eqnarray} \label{NEWPROJECTION}
 {\mathcal{P}} \rho &=&
 {\mathrm{tr}}_E \left\{ \Pi_1 \rho \right\} \otimes \frac{1}{N_1} \Pi_1
 + {\mathrm{tr}}_E \left\{ \Pi_2 \rho \right\} \otimes \frac{1}{N_2}
 \Pi_2 \nonumber \\
 &\equiv& \rho_S^{(1)} \otimes \frac{1}{N_1} \Pi_1
 + \rho_S^{(2)} \otimes \frac{1}{N_2} \Pi_2.
\end{eqnarray}
This projection belongs to the class of superoperators introduced
in Eq.~(\ref{PROJ-NEW}). By this ansatz the total system's state
is approximated by a separable but non-factorizing state. The
dynamical variables are the un-normalized density matrices
$\rho_S^{(1)}$ and $\rho_S^{(2)}$ which are correlated with the
projections onto the lower and the upper band, respectively. The
reduced density matrix of the two-state system is found with the
help of Eq.~(\ref{RED-DENSITY}):
\begin{equation} \label{REDUCED}
 \rho_S(t) = {\mathrm{tr}}_E \{ {\mathcal{P}} \rho(t) \}
 = \rho_S^{(1)}(t) + \rho_S^{(2)}(t).
\end{equation}
Assuming that the initial state is of the correlated form
\begin{equation}
 \rho(0) = \rho_S^{(1)}(0) \otimes \frac{1}{N_1} \Pi_1
 + \rho_S^{(2)}(0) \otimes \frac{1}{N_2} \Pi_2,
\end{equation}
we have ${\mathcal{P}}\rho(0)=\rho(0)$ and the inhomogeneous term
of the TCL equation (\ref{TCL-MASTER}) vanishes.

\subsubsection{The second-order master equation}
Using the new projection superoperator (\ref{NEWPROJECTION}) we
get the following second-order TCL equation:
\begin{eqnarray} \label{DERIV1}
 \frac{d}{dt} {\mathcal{P}}\rho(t) &=& \dot{\rho}_S^{(1)}(t) \otimes
 \frac{1}{N_1} \Pi_1 + \dot{\rho}_S^{(2)}(t) \otimes
 \frac{1}{N_2} \Pi_2 \nonumber \\
 &=& {\mathcal{K}}_2(t) {\mathcal{P}}\rho(t),
\end{eqnarray}
where the TCL generator takes the form:
\begin{eqnarray} \label{DERIV2}
 \lefteqn{ {\mathcal{K}}_2(t) {\mathcal{P}}\rho(t) = } \\
 && \int_0^t dt_1 h(t-t_1) \nonumber \\
 && \times \left[ 2\gamma_1 \sigma_+ \rho_S^{(2)} \sigma_-
 - \gamma_2 \{ \sigma_+\sigma_-,\rho_S^{(1)}\}
 \right]  \otimes \frac{1}{N_1} \Pi_1 \nonumber \\
 && + \int_0^t dt_1 h(t-t_1) \nonumber \\
 && \times \left[ 2\gamma_2 \sigma_- \rho_S^{(1)} \sigma_+
 - \gamma_1 \{ \sigma_-\sigma_+,\rho_S^{(2)}\}
 \right]  \otimes \frac{1}{N_2} \Pi_2. \nonumber
\end{eqnarray}
Combining Eqs.~(\ref{DERIV1}) and (\ref{DERIV2}) and assuming
again $\delta\varepsilon \cdot t \gg 1$ we find the equations of
motion:
\begin{eqnarray}
 \frac{d}{dt} \rho_S^{(1)}(t) &=&
 \gamma_1 \sigma_+ \rho_S^{(2)} \sigma_-
 -\frac{\gamma_2}{2} \{ \sigma_+\sigma_-, \rho_S^{(1)} \},
 \label{MASTER1} \\
 \frac{d}{dt} \rho_S^{(2)}(t) &=&
 \gamma_2 \sigma_- \rho_S^{(1)} \sigma_+
 -\frac{\gamma_1}{2}\{ \sigma_-\sigma_+, \rho_S^{(2)} \}.
 \label{MASTER2}
\end{eqnarray}
This is a coupled system of first-order differential equations for
the two density matrices $\rho_S^{(1)}(t)$ and $\rho_S^{(2)}(t)$.
The elements of these matrices are written as
\begin{equation}
 \rho^{(1)}_{ij}(t) = \langle i |\rho_S^{(1)}(t)|j\rangle, \qquad
 \rho^{(2)}_{ij}(t) = \langle i |\rho_S^{(2)}(t)|j\rangle.
\end{equation}

\begin{figure}[htb]
\includegraphics[width=0.95\linewidth]{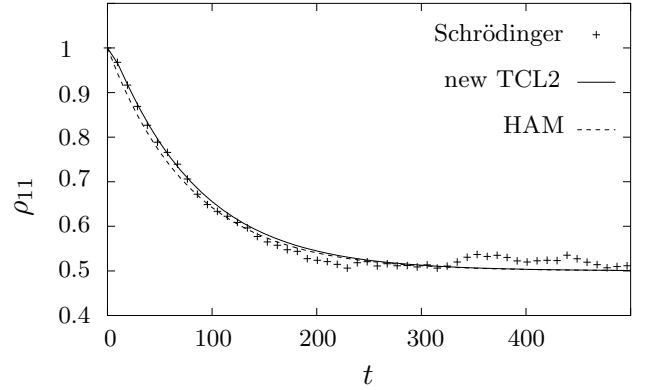}
\caption{comparison of the second-order TCL approximation using
the new projection superoperator [Eq.~(\ref{RHO11-TCL2})], of the
approximation given by HAM [Eq.~(\ref{HAM2})], and of the
numerical solution of the Schr\"odinger equation. Parameters:
$N_1=N_2=500$, $\delta\varepsilon=0.5$, and
$\lambda=0.001$.\label{TCL-fig1}}
\end{figure}

The equations (\ref{MASTER1}) and (\ref{MASTER2}) can now be used
to derive an equation of motion for the reduced density matrix,
making use of Eq.~(\ref{REDUCED}). First, we get from
Eq.~(\ref{MASTER2}):
\begin{eqnarray}
 \frac{d}{dt} \rho^{(2)}_{11}(t) &=& 0, \label{ABL-1} \\
 \frac{d}{dt} \rho^{(2)}_{00}(t) &=& \gamma_2\rho_{11}^{(1)}(t)
 - \gamma_1 \rho_{00}^{(2)}(t). \label{ABL-2}
\end{eqnarray}
We assume again that initially only the lower band is populated:
\begin{equation} \label{ABL-3}
 \rho_S^{(2)}(0) = 0.
\end{equation}
It thus follows from Eq.~(\ref{ABL-1}) that
\begin{equation} \label{ABL-5}
 \rho_{11}^{(2)}(t) \equiv 0.
\end{equation}
From Eq.~(\ref{MASTER1}) we find
\begin{equation} \label{ABL-4}
 \frac{d}{dt} \rho^{(1)}_{11}(t) = \gamma_1\rho_{00}^{(2)}(t)
 - \gamma_2 \rho_{11}^{(1)}(t).
\end{equation}
From Eqs.~(\ref{ABL-4}) and (\ref{ABL-2}) we see that the quantity
$\rho^{(1)}_{11}(t) + \rho^{(2)}_{00}(t)$ is constant. With the
help of the initial condition (\ref{ABL-3}) we thus have
\begin{equation}
 \rho_{00}^{(2)}(t) =  \rho^{(1)}_{11}(0) - \rho^{(1)}_{11}(t).
\end{equation}
Substituting this into Eq.~(\ref{ABL-4}) we find:
\begin{equation} \label{ABL-6}
 \frac{d}{dt} \rho^{(1)}_{11}(t) = -(\gamma_1+\gamma_2)\rho_{11}^{(1)}(t)
 + \gamma_1 \rho_{11}^{(1)}(0).
\end{equation}
Since $\rho_{11}(t) \equiv \rho_{11}^{(1)}(t)$ because of
Eq.~(\ref{ABL-5}), we finally arrive at the equation of motion for
the populations:
\begin{equation} \label{POP}
 \frac{d}{dt} \rho_{11}(t) = -(\gamma_1+\gamma_2) \rho_{11}(t)
 + \gamma_1 \rho_{11}(0).
\end{equation}
In a similar manner one is led to the equation of motion for the
coherences:
\begin{equation} \label{COH}
 \frac{d}{dt} \rho_{01}(t) = -\frac{\gamma_2}{2} \rho_{01}(t).
\end{equation}
The dynamics of the reduced density matrix is thus determined by
Eqs.~(\ref{POP}) and (\ref{COH}). These are time-local first-order
differential equations with constant coefficients. They are
identical to the equations of motion obtained using HAM. In
particular, the solution of Eq.~(\ref{POP}) is given by the
expression (\ref{HAM2}). Hence, we conclude that the lowest order
of the TCL expansion with the projection superoperator introduced
in Eq.~(\ref{NEWPROJECTION}) indeed reproduces the HAM prediction.

\begin{figure}[htb]
\includegraphics[width=0.95\linewidth]{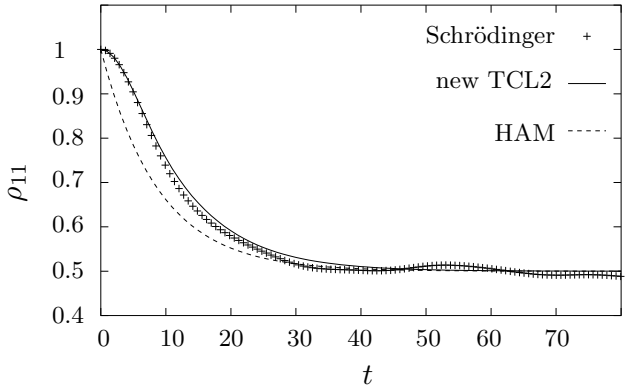}
\caption{the same as Fig.~\ref{TCL-fig1} for
$\lambda=0.003$.\label{TCL-fig2}}
\end{figure}

\begin{figure}[htb]
\includegraphics[width=0.95\linewidth]{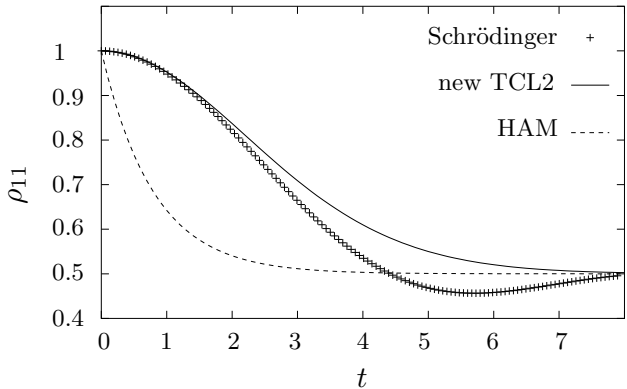}
\caption{the same as Fig.~\ref{TCL-fig1} for
$\lambda=0.01$.\label{TCL-fig3}}
\end{figure}

We observe that the dynamics of the populations $\rho_{11}(t)$ is
strongly non-Markovian because of the presence of the initial
condition $\rho_{11}(0)$ on the right-hand side of
Eq.~(\ref{POP}). This term expresses a pronounced memory effect,
namely it implies that the dynamics of the populations never
forgets its initial data. Note also that the dynamics of the
reduced density matrix is not in Lindblad form and does even not
represent a semigroup. It does, however, lead to a positive
dynamical map, as can easily be verified.

In the transition from Eq.~(\ref{DERIV2}) to the
Eqs.~(\ref{MASTER1}) and (\ref{MASTER2}) we have assumed for
simplicity that the times $t$ considered satisfy the condition
$\delta\varepsilon\cdot t \gg 1$. Without this condition the
master equations (\ref{MASTER1}) and (\ref{MASTER2}) must be
replaced by
\begin{eqnarray}
 \frac{d}{dt} \rho_S^{(1)}(t) &=&
 \int_0^t dt_1 h(t-t_1) \label{MASTER1-full} \\
 && \times \left[2\gamma_1 \sigma_+ \rho_S^{(2)}(t) \sigma_-
 -\gamma_2\{ \sigma_+\sigma_-, \rho_S^{(1)}(t) \}\right],
  \nonumber \\
 \frac{d}{dt} \rho_S^{(2)}(t) &=&
 \int_0^t dt_1 h(t-t_1) \label{MASTER2-full} \\
 && \times \left[2\gamma_2 \sigma_- \rho_S^{(1)}(t) \sigma_+
 -\gamma_1\{ \sigma_-\sigma_+, \rho_S^{(2)}(t) \}\right]. \nonumber
\end{eqnarray}
These equations describe the full time dependence as it is
predicted by the TCL expansion in second order. They lead to the
following populations of the upper level:
\begin{equation} \label{RHO11-TCL2}
 \rho_{11}(t) = \rho_{11}(0) \left[ \frac{\gamma_1}{\gamma_1+\gamma_2}
 + \frac{\gamma_2}{\gamma_1+\gamma_2}
 e^{-\Gamma(t)} \right],
\end{equation}
where
\begin{equation}
 \Gamma(t) = 2(\gamma_1+\gamma_2)\int_0^t dt_1 \int_0^{t_1} dt_2
 h(t_1-t_2),
\end{equation}
and the function $h(\tau)$ is given by Eq.~(\ref{h-def}).

In Figs.~\ref{TCL-fig1}, \ref{TCL-fig2}, and \ref{TCL-fig3} we
compare the result given by Eq.~(\ref{RHO11-TCL2}) with the
prediction of HAM and with numerical simulations of the
Schr\"odinger equation. The figures clearly show that already the
lowest order of the TCL expansion with the new projection
superoperator gives a good approximation of the dynamics. It not
only yields the correct stationary state, but also reasonable
predictions on the relaxation times even for rather strong
couplings, where deviations from the HAM result are large. Note
that the rates $\gamma_{1,2}$ of Fig.~\ref{TCL-fig1} differ from
those of Fig.~\ref{TCL-fig3} by a factor of $100$, and that the
parameters of Fig.~\ref{TCL-fig3} correspond to the ratio
$\gamma_{1,2}/\delta\varepsilon \approx 1.3$.

\subsubsection{The master equation of fourth order}
We have seen that the lowest order of the TCL expansion obtained
with the help of the correlated projection superoperator is
capable of reproducing the prediction of the HAM approximation,
and even improves this approximation for larger couplings. The
question is now: What happens in higher orders of the expansion?
As our numerical simulations indicate, higher-order corrections
should be small for all times. We show that, by contrast to the
case of the standard projection, this is indeed the case.

The fourth-order contribution to the TCL generator is obtained by
using the projection superoperator defined by
Eq.~(\ref{NEWPROJECTION}) in the general expression (\ref{K4}).
One finds:
\begin{widetext}
\begin{eqnarray*}
{\mathcal{K}}_4(t) {\mathcal{P}}\rho
 &=& \int_0^t dt_1 \int_0^{t_1} dt_2 \int_0^{t_2} dt_3 \\
 &~& \;\;\; \left[ 2(\gamma_1^2+\gamma_1\gamma_2)
 \left[h(t-t_2)h(t_1-t_3)+h(t-t_3)h(t_1-t_2)\right]
 \sigma_+ \rho_S^{(2)} \sigma_- \right. \\
 &~& \;\;\; -2\gamma_1\gamma_2
 \left[h(t-t_2)h(t_1-t_3)+2h(t-t_3)h(t_1-t_2)\right]
 \sigma_+\sigma_- \rho_S^{(1)} \sigma_+\sigma_- \\
 &~& \;\;\; - \left.
 \left[\gamma_2^2h(t-t_2)h(t_1-t_3)+(\gamma_2^2-\gamma_1\gamma_2)h(t-t_3)h(t_1-t_2)\right]
 \{\sigma_+ \sigma_-,\rho_S^{(1)} \} \right]
 \otimes \frac{1}{N_1} \Pi_1 \\
 &~& + \left[ 2(\gamma_2^2+\gamma_1\gamma_2)
 \left[h(t-t_2)h(t_1-t_3)+h(t-t_3)h(t_1-t_2)\right]
 \sigma_- \rho_S^{(1)} \sigma_+ \right. \\
 &~& \;\;\; -2\gamma_1\gamma_2
 \left[h(t-t_2)h(t_1-t_3)+2h(t-t_3)h(t_1-t_2)\right]
 \sigma_-\sigma_+ \rho_S^{(2)} \sigma_-\sigma_+ \\
 &~& \;\;\; - \left.
 \left[\gamma_1^2h(t-t_2)h(t_1-t_3)+(\gamma_1^2-\gamma_1\gamma_2)h(t-t_3)h(t_1-t_2)\right]
 \{\sigma_- \sigma_+,\rho_S^{(2)} \} \right]
 \otimes \frac{1}{N_2} \Pi_2.
\end{eqnarray*}
\end{widetext}
Here, $h(\tau)$ is again a function which is sharply peaked at
$\tau=0$ and may be replaced by the delta function $\delta(\tau)$
for times which are large compared to the inverse band width. The
decisive point is the following. When carrying out the three-fold
time integrations in the above expression, no terms emerge which
grow like powers of $t$. This is due to the fact that the
integrands do not contain terms of the form $h(t_2-t_3)$. As a
result, the contribution from ${\mathcal{K}}_4(t)$ remains small
for all times $t$, and the limit $t \rightarrow \infty$ of the
generator exists.

To illustrate this point let us model the function $h(\tau)$ by
\begin{equation}
 h(\tau) = \frac{\delta\varepsilon}{2}
 e^{-\delta\varepsilon\cdot|\tau|},
\end{equation}
which approaches $\delta(\tau)$ for infinite band width. The
time-integrations may then easily be carried out to give:
\begin{eqnarray*}
 \int_0^t dt_1 \int_0^{t_1} dt_2 \int_0^{t_2} dt_3 h(t-t_2)h(t_1-t_3)
 &\approx& \frac{1}{8\delta\varepsilon}, \\
 \int_0^t dt_1 \int_0^{t_1} dt_2 \int_0^{t_2} dt_3 h(t-t_3)h(t_1-t_2)
 &\approx& \frac{1}{8\delta\varepsilon}.
\end{eqnarray*}
Hence we see that ${\mathcal{K}}_4(t)$ becomes time-independent
for $\delta\varepsilon\cdot t \gg 1$ and that the fourth order of
the expansion leads to the equations of motion:
\begin{eqnarray}
 \frac{d}{dt} \rho_S^{(1)}(t) &=&
 \Gamma_1 \sigma_+ \rho_S^{(2)} \sigma_-
 -\frac{\Gamma_2}{2} \{ \sigma_+\sigma_-, \rho_S^{(1)} \} \nonumber \\
 &~& -\Gamma_3 \sigma_+\sigma_- \rho_S^{(1)} \sigma_+\sigma_-, \\
 \frac{d}{dt} \rho_S^{(2)}(t) &=&
 \tilde{\Gamma}_2 \sigma_- \rho_S^{(1)} \sigma_+
 -\frac{\tilde{\Gamma}_1}{2} \{ \sigma_-\sigma_+, \rho_S^{(2)} \} \nonumber \\
 &~& -\tilde{\Gamma}_3 \sigma_-\sigma_+ \rho_S^{(2)} \sigma_-\sigma_+,
\end{eqnarray}
where we have introduced the rates:
\begin{eqnarray}
 \Gamma_1 &=& \gamma_1 \left[ 1+\frac{\gamma_1+\gamma_2}{2\delta\varepsilon}
 \right], \\
 \Gamma_2 &=& \gamma_2 \left[ 1+\frac{2\gamma_2-\gamma_1}{4\delta\varepsilon}
 \right], \\
 \Gamma_3 &=& \tilde{\Gamma}_3 =
 \frac{3\gamma_1\gamma_2}{4\delta\varepsilon}, \\
 \tilde{\Gamma}_1 &=& \gamma_1 \left[ 1+\frac{2\gamma_1-\gamma_2}{4\delta\varepsilon}
 \right], \\
 \tilde{\Gamma}_2 &=& \gamma_2 \left[ 1+\frac{\gamma_1+\gamma_2}{2\delta\varepsilon}
 \right].
\end{eqnarray}
This shows that the fourth order merely yields corrections of
order ${\mathcal{O}}(\gamma_{1,2}/\delta\varepsilon)$ to the
equations of motion found in second order. Thus, as expected the
TCL series obtained with the correlated projection superoperator
indeed leads to a systematic perturbation expansion around the
approximation suggested by HAM.

We finally mention that the equations of motion for the
populations of the reduced system in fourth order take the form:
\begin{equation}
 \frac{d}{dt} \rho_{11}(t) = -(\Gamma_1+\tilde{\Gamma}_2) \rho_{11}(t)
 + \Gamma_1 \rho_{11}(0),
\end{equation}
which is easily solved to yield:
\begin{equation}
 \rho_{11}(t) = \rho_{11}(0) \left[ \frac{\gamma_1}{\gamma_1+\gamma_2}
 + \frac{\gamma_2}{\gamma_1+\gamma_2}
 e^{-(\Gamma_1+\tilde{\Gamma}_2)t} \right].
\end{equation}
The interesting point to note is that the stationary state is
identical to the one found in second order. Thus, the stationary
state is not affected by the fourth-order corrections. However,
the rate of the relaxation into the stationary state is found to
be
\begin{equation}
 \Gamma_1+\tilde{\Gamma}_2 = (\gamma_1+\gamma_2)
 \left[1+\frac{\gamma_1+\gamma_2}{2\delta\varepsilon}\right],
\end{equation}
which is seen to be larger than the relaxation obtained in second
order.

\section{Conclusions}\label{CONCLU}
In this paper we have analyzed non-Markovian quantum processes by
means of the time-convolutionless projection operator technique
and of the Hilbert space average method. It has been demonstrated
that by use of a class of projection operators which project onto
correlated system-environment states, an efficient
non-perturbative treatment of certain non-Markovian processes is
possible. The correlated projections have been shown to correspond
to the idea of a best guess underlying the approximation of HAM.

The general mathematical formalism of the projection operator
techniques does not tell us which projection superoperator should
be used for a given system-environment model. The choice of an
appropriate projection ${\mathcal{P}}$ depends on the structure of
the model under study and must be based on physical
considerations. The aim is of course an {\textit{efficient}}
description, i.~e., a description which can be expected to yield
accurate results even in low orders of the coupling. Once a
certain projection ${\mathcal{P}}$ has been chosen, one can use
the perturbation expansion in order to check explicitly whether or
not higher orders remain small and, thus, whether or not
${\mathcal{P}}$ enables a computationally efficient treatment of
the reduced dynamics.

We emphasize that the projection operator techniques yield an
expansion of the equations of motion for the relevant variables
${\mathcal{P}}\rho(t)$, and {\textit{not}} an expansion of these
variables itself. Different projection superoperators lead to
different sets of relevant variables and, hence, to equations of
motion with completely different structures. Consequently, the
usage of different projections implies a complete re-organization
of the perturbation expansion. This point has been demonstrated
here by means of a specific system-environment model. As we have
seen, the TCL technique which is based on the correlated
projection superoperator yields accurate results for this model
already in lowest order, while the standard TCL procedure fails in
any finite order of the coupling.

Our results suggest applications to other models showing strong
non-Markovian effects. For example, it is clear that the class of
correlated projections introduced here may also be applied to a
generalization of the model studied in Sec.~\ref{Sec-Appl} which
involves any number of well-separated environmental energy bands.

In Sec.~\ref{Sec-Superoperators} we have formulated two general
conditions for suitable superoperators ${\mathcal{P}}$. The first
one [Eq.~(\ref{PROJECTION})] is simply the condition that the map
${\mathcal{P}}$ be a projection operator. The second condition
[Eq.~(\ref{CONSISTENCY})] requires that the projection
${\mathcal{P}}\rho$ of any state $\rho$ of the composite system
contains the full information which is necessary to extract the
reduced density matrix $\rho_S$ from the relevant variables. It is
easy to construct classes of superoperators which satisfy these
two conditions and are even more general than the class of
correlated projections investigated here. As mentioned already the
latter project the total state onto a separable, classically
correlated system-environment state. This suggests exploiting the
possibility of using superoperators which project onto
nonseparable, entangled quantum states. In this way one might be
able to investigate the dynamical significance of entanglement in
non-Markovian quantum processes.

\begin{acknowledgments}
We thank G. Mahler for fruitful discussions on this subject.
Financial support for M. Michel and J. Gemmer by the Deutsche
Forschungsgemeinschaft is gratefully acknowledged.
\end{acknowledgments}

\end{document}